\documentclass[12pt]{article}
\usepackage{tikz}
\usepackage{amsmath}
\usepackage{mathtools}
\usepackage{fancyhdr}
\usepackage{gensymb}
\usepackage{graphicx}
\usepackage{cite}
\usetikzlibrary{decorations.pathmorphing, arrows, patterns}
\tikzset{snake it/.style={-stealth,
decoration={snake, 
    amplitude = .4mm,
    segment length = 2mm,
    post length=0.9mm},decorate}}
\usepackage{amssymb}

\newcommand{\midarrow}{\tikz \draw[-triangle 90] (0,0) -- +(.1,0);}

\usepackage[a4paper, total={6in, 10in}]{geometry}

\begin{document}
\title{\Large{\textbf{Covariant dynamics on the energy-momentum space: scalar field theory}}}
\date{}
\author{Boris Iveti\'c
\footnote{bivetic@yahoo.com}
  \\Wöhlergasse 6, 1100 Vienna, Austria \\}

\maketitle

\begin{abstract}
A scalar field theory is constructed on an energy-momentum background of constant curvature. The generalization of the usual Feynamn rules for the flat geometry follows from the requirement of their covariance. The main result is that the invariant amplitudes are finite at all orders of the perturbation theory, due to the finitness of the momentum space. Finally, the relation with a field theory in spacetime representation is briefly discussed. 
\end{abstract}

Ultraviolet divergencies of various quantum field theories (QFTs) all have a common source, that is the integration over infinite momenta. If a suitable cut-off scale is introduced, the divergencies disappear. In the context of effective field theories, the introduction of such a scale is physically justified, as this scale represents the upper bound of the validity of an effective theory. For a fundamental theory this approach is not favorable, since a finite scale requires a physical interpretation, and letting it go to infinity at the end of calculation does not always guarantee a finite result. The cut-off approach has therefore largely been abandoned, and was subsequently replaced by an extensive methodology of dealing with the infinities through dimensional regularization and renormalization group. This program, initiated by Stueckelberg and Petermann \cite{stu}, and Gell-Mann and Low \cite{gm}, and subsequently formalized  through the works of, among others, Weinberg \cite{wein}, t'Hooft and Veltman \cite{tho}, and  Wilson \cite{wil},  does indeed offer a methodical way of subtracting the infinities appearing at different orders of perturbation theory, leaving the physical, finite contributions that are readily checked experimentally, upon a suitable numerical input of a small number of parameters of the theory. This makes QFT one of the most succesfull physical theories.

Its phenomenological success notwithstanding, the criticism of the renormalization methodology of handling the infinities still persists, in particular among more mathematically oriented physicists, as well as among pure mathematicians.. One way to avoid this criticism of the renormalization is to try to avoid divergent quantities alltogether,  by going back to the cut-off scheme. But how can the cut-off scale be physically justified, and the corresponding QFT at the same time be fundamental? The answer could be provided by the geometry of the momentum space.

It is usually assumed that the momentum background, on which certain QFT is defined, is flat, despite the fact that there is no explicit requirement to be so from the axioms of the theory. This is presumably done out of simplicity. Flat geometry is the simplest one, and the only one that does not require parameters to describe it. More complicated geometries do: the geometry of a sphere requires one parameter (radius), the geometry of an ellipse at least two, and so on. 

In this work we shall deal with the simplest geometrical deviation from the flat case, i.e. that of the (pseudo)-spherical energy-momentum space. The radius of its curvature will serve as a cut-off scale. This idea was introduced by Snyder \cite{sny} in 1947., precisely to deal with the infinities in the then emerging theory of quantum fields. Formulation of QFT on maximally symmetric spaces was investigated  in the following decdes \cite{gol, kady1, kady2, mir1, mir2}, but the real upsurge in the interest in the Snyder model came through research in quantum gravity and string theory \cite{qg1,qg2} and with the development of quantum groups \cite{qg3}. Of the recent efforts to formulate QFT on Snyder space, we point out to \cite{bm,gl,rv,fm}. The formulation we propose here is novel, and is based on using only momentum space geometrical invariants for describing the dynamics. It represents a conceptual follow-up to the study of dynamics in the context of the classical electrodynamics and the nonrelativistic quantum physics on Snyder space, that were dealt with in \cite{bor} and \cite{bor2}, respectively. 
 \\ \\
\textbf{Axioms}\\
We start by listing the definitions of and the demands on an energy-momentum background on which a physical field theory is to be defined:
\begin{itemize}
\item[1.]  Energy-momentum manifold $\mathcal P$ is a four-dimensional  Riemmanian manifold described (locally) by a metric $g_{\mu\nu}(p)$ or (globally) by a distance function $d(p,q)$, which is a geodesic distance between the points $p$ and $q$ on the manifold.\\
\item[2.] Manifold $\mathcal P$ contains a distinguished unique point $p_V$ called the origin, which is identified with the absolute vacuum. The coordinates of the vacuum are in any reference frame (any system of coordinates) $p_{V}=(0,0,0,0)$. In addition to the origin, manifold $\mathcal P$ contains the set of points labeled the points at infinity, which is a set of points with a maximal (finite or infinite) distance from the origin. \footnote{We include points at infinity to the manifold from technical reasons (see axiom 5 below). This does not affect the dynamics of finite points in any way.}\\
\item[3.] Elementary particles are the energy-momentum excitations with respect to the vacuum. Mass of an elementary particle on the point $p$ on $\mathcal P$ is defined as its geodesic distance from the origin, $$m(p)=d(0,p).$$\\
\item[4.] Manifold $\mathcal P$ must be able to accomodate a group of isometries that leave the origin invariant.\\
\item[5.] Manifold $\mathcal P$ must be able to accomodate a group of isometries that leave the points at infinity invariant. 
\end{itemize}

In a very broad sense of the word we call them the axioms of the momentum space. They represent a minimal set of physically reasonable requirements for a momentum backgrounds, in a sense that relaxing any of them would lead to theories with much different physical concepts that the ones we are used to.\footnote{A less restrictive set of axioms was proposed in \cite{relloc1}.} For instance,  the group in 4 is the usual Lorentz group, and its physical origin lies in the demand that all the observers agree on the value of the mass parameter (postion of the origin). The froup in 5. is the group of translations or displacements, and it defines the energy-momentum conservation law (the momenta addition) in the interaction of elementary particles.\footnote{The group parameters in this case are the points of $\mathcal P$ themselves, and the action of the group is symbolically presented with $\oplus$, with the understanding that $k\oplus p$ represents the displacement \textit{of} the point $p$ \textit{by} the point $k$.} Physically the demand of displacement isometry arises out of the demand that no two particles $p$ and $q$ should be able to reduce or increase their mutual distance upon absorbing/emitting particles of the same energy-momentum $k$, i.e. $d(k\oplus p, k\oplus q)=d(p,q)$. The demand for the invariance of the points at infinity reflects a physical demand that no finite number of finite displacements of some point that does not belong to the set of points at infinity could displace it to the point at infinity.

We note some important direct consequences of the above axioms. 

In order to accomodate enough global isometries, manifold $\mathcal P$ must necessarily be of a maximally symmetric type, with either positive, negative or vanishing curvature. Under the assumption that the radius of curvature is very large, for the points $p$ that lie close to the origin the manifold looks flat. It is in this sense that our current theory of the Minkowskian $\mathcal P$ can be understood as a low energy approximation of a more fundamental theory. Additionally, this gives a very important consistency condition in the construction of a fundamental theory, namely, at any instance one should be able to reproduce the standard theory by setting the curvature to zero. 

Furthermore, no explicit mention of any prefered system of coordinates implies that the dynamics must be defined in terms of geometrical invariants only. Otherwise, one would need to add to the set of axioms a rule which defines absolute physical quantities of energy-momentum in terms of some specific coordinates of $\mathcal P$. In particular, 3., which was given earlier in \cite{relloc1}, represents geometrization of the usual dispersion relation $p^2=m^2$. It is only for the flat space that the points of the manifold are at the same time vectors (the radii-vectors), and $p^2=\eta^{\mu\nu}p_\mu p_\nu$ consequently scalars. The latter is then naturally generalized for non-flat spaces to the square of the distance function from the origin.

Having established that the energy-momentum background is maximally symmetric, we review some of the mathematical details that we will need in the following, that have also previously been reported elsewhere.
\\ \\
\textbf{Mathematical prerequisites}\\
A maximally symmetric energy-momentum space is usually realised as a 4-surface emebedded in a 5-dimensional Minkowskian space,

\begin{equation}\label{emb}
\eta_0^2-\eta_1^2-\eta_2^2-\eta_3^2-\eta_4^2=\mathcal P^2
\end{equation}
where $\mathcal P$ is the radius, which we in the following take as real and finite.

Physical energies and momenta are given as some function of the emebedding space coordinates,

\begin{equation}
\eta_\mu=h(p^2/\mathcal P^2)p_\mu, \ \ \ \ \eta_4=\mathcal P\sqrt{1+h^2p^2/\mathcal P^2 }
\end{equation}
 The metric in coordinates $p$ is 

\begin{equation}
g_{\mu\nu}(p)=h^2\eta_{\mu\nu}+\frac{4h'(h+h'p^2/\mathcal P^2)-h^4}{\mathcal P^2+h^2p^2}p_\mu p_\nu,
\end{equation}
where $h'\equiv\partial h/ \partial (p^2/\mathcal P^2)$, and the distance function is
\begin{equation}
d(p,k)=\mathcal P \text{Arcosh} \left( h_ph_k(pk)/\mathcal P^2-\sqrt{1+h_p^2p^2/\mathcal P^2}\sqrt{1+h_k^2k^2/\mathcal P^2}      \right),
\end{equation}
where $h_p=h(p^2/\mathcal P^2)$, $h_k=h(k^2/\mathcal P^2)$. The law of the momenta addition, or the energy-momentum conservation law is given by 

\begin{equation}
p_\mu\oplus k_\mu=h^{-1}\left( \mathcal P^{-2}\mathcal K^2  \right)\mathcal K_\mu,
\end{equation} 
where

\begin{equation}
\mathcal K_\mu=h_pp_\mu+h_kk_\mu\left( \sqrt{1+\mathcal P^{-2}h_p^2p^2}+\frac{\mathcal P^{-2}h_ph_k(pk)}{1+\sqrt{1+\mathcal P^{-2}h_k^2k^2}}   \right).
\end{equation} 

The enrgy-momentum integral is changed accordingly,

\begin{equation}\label{int}
\int d^4p\to\int d\Omega_p=\int \sqrt{det g}d^4p
\end{equation}

Finally, the delta function generalizes to

\begin{equation}
\delta(p-k)\to \delta(p\ominus k),
\end{equation}

wchich is defined in the distributional sense as

\begin{equation}
\int d\Omega_p f(p) \delta(p\ominus k)=f(k).
\end{equation}
\textbf{Feynman rules}\\
The set of Feynman rules represent a graphical metod of computing invariant amplitudes of scattering cross sections perturbatively in the coupling constant. They are usually first derived on the spacetime, after which they are Fourier transformed on the energy-momentum space. Once this is accomplished, the momentum space Feynman rules represent a full and self-consistent sets of rules for calculating the scattering ampitudes. Here we shall use the rules on the energy-momentum background as our starting point, and write them covariantly, i.e. in the form applicable to any geometry or coordinatization of the energy-momentum space.  The idea is to maintain the same field theory as in the canonical case, only applicapbe to the general background geometry. For the sake of concretness, we choose scalar field theory, with all our considerations directly generalizable to the theories involving vector and spinor fields. In the standard case $\phi^n$ theory is given by the Lagrangian:

\begin{equation}
\mathcal L=\phi(p^2-m^2)\phi-\frac{g}{n!}\phi^n
\end{equation}

On a general energy-momentum background this is generalized to

\begin{equation}\label{lag}
\mathcal L=\phi(d^2(p,0)-m^2)\phi-\frac{g}{n!}\phi^n
\end{equation}
in order to accomodate for a generalization in the dispersion relation given in axiom 4 above.

The set of usual Feynman rules can be found in any textbook on field theory. Internal lines (those not connected to external points) get free propagators $$D_F(p)=(p^2-m^2-i\epsilon)^{-1}.$$ Vertices come from interactions in the Lagrangian. They get factors of the
coupling constant times $i$, with momentum conservation imposed at each vertex via a delta function $$\delta(p_{out}-p_{in}).$$
Finally, one integrates over undetermined 4-momenta, $$\int d^4p,$$
and sums over all possible diagrams.

This is readily generalized to an arbitrary geometry and coordinatizetion of the momentum space. The Klein-Gordon equation following from free part of (\ref{lag}) is
\begin{equation}
(d^2(p,0)-m^2)\phi(p)=0,
\end{equation}
and the free particle propagator is, as in the standard case, Green's function of the Klein-Gotdon equation,

\begin{equation}
(d^2(p,0)-m^2)D_F(p)=-1,
\end{equation}
whose solution (in the distributional sense) is, according to the Sokhotski–Plemelj theorem,

\begin{equation}
D_F(p)=\frac{1}{d^2(p,0)-m^2\pm i\epsilon},
\end{equation}
with the $+i\epsilon$ a choice consistent with causality, as in the canonical case. Vertex factors remain the same, with the momentum conservation imposed via generalized delta function, 

\begin{equation}
\delta(p_{out}\ominus p_{in})
\end{equation}
and the integration over undefined momenta proceeds according to (\ref{int}). 

For instance, let us take $\phi^3$ theory at the tree-level. In the $s$-channel the diagram is,

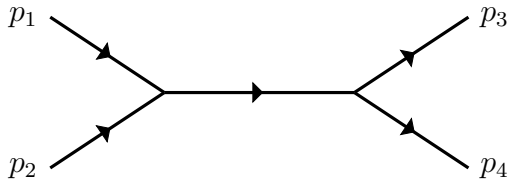
\begin{figure}[h]
\begin{center}
\begin{tikzpicture}[thick]
\begin{scope}[very thick, every node/.style={sloped,allow upside down}]  
\draw (-5,1) -- node{\midarrow} (-3.5,0);  
\draw   (-5,-1) -- node{\midarrow} (-3.5,0);
\draw   (-3.5,0) -- node{\midarrow} (-1,0);
\draw (-5.7,1) node[anchor= west] {$ p_1 $};
\draw (-5.7,-1) node[anchor= west] {$ p_2 $};
\draw   (-1,0) -- node{\midarrow} (0.5,1);
\draw  (-1,0) --node{\midarrow} (0.5,-1);
\draw (0.5,1) node[anchor= west] {$ p_3 $};
\draw (0.5,-1) node[anchor= west] {$ p_4 $};
\end{scope} 
\end{tikzpicture}
\end{center}
\end{figure}
and its contributions to the amplitude is

\begin{equation}
i\mathcal M_s=\frac{(ig)^2}{d^2(p_1\oplus p_2,0)-m^2+i\epsilon}.
\end{equation}
Despite the fact that the law of momenta addition is non-commutative, $p_1\oplus p_2\neq p_2\oplus p_1$, the amplitude is independent of which particle we call one and which two. This is due to the fact that

\begin{equation}
d(p\oplus k,0)=d(k\oplus p,0).
\end{equation}
Similarly, in $t$- and $u$-channels,   

\begin{equation}
i\mathcal M_t=\frac{(ig)^2}{d^2(p_1\ominus p_3,0)-m^2+i\epsilon}, \ \ \ i\mathcal M_u=\frac{(ig)^2}{d^2(p_1\ominus p_4,0)-m^2+i\epsilon},
\end{equation}
where $p_i\ominus p_j=p_i\oplus(-p_j)$.

What about the diagrams containing loops? Let us take for example the leading correction to the free propagator in $\phi^3$ theory: 

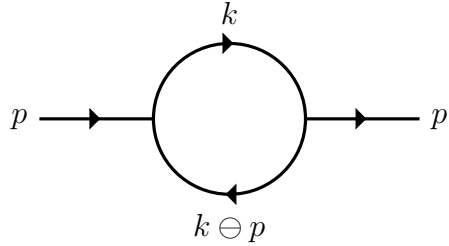
\begin{figure}[h]
\begin{center}
\begin{tikzpicture}[thick]
\begin{scope}[very thick, every node/.style={sloped,allow upside down}]  
\draw   (-0.5,0) -- node{\midarrow} (1,0);
\draw   (-4,0) -- node{\midarrow} (-2.5,0);
\draw (-1.5,0) circle (1);
\draw (-1.5,1) node{\midarrow};
\draw (-1.5,-1) -- node{\midarrow}(-1.501,-1);
\draw (-4,0) node[anchor= east] {$ p $};
\draw (1,0) node[anchor= west] {$ p $};
\draw (-1.5,1.1) node[anchor= south] {$ k $};
\draw (-1.5,-1.1) node[anchor= north] {$ k\ominus p $};
\end{scope} 
\end{tikzpicture}
\end{center}
\end{figure}
According to our rules, it is

\begin{equation}
i\mathcal M=-\frac{g^2}{2}\int \frac{d\Omega_k}{(2\pi)^4}\frac{i}{d^2(k\ominus p,0)-m^2+i\epsilon} \frac{i}{d^2( k,0)-m^2+i\epsilon}. 
\end{equation}
This is manifestly finite, due to the finite volume of the momentum space. Likewise, a diagram with an arbitrary number of loops, whose generic form is 
\begin{equation}
i\mathcal M=\prod_i\int \frac{d\Omega_{p_i}}{(2\pi)^4}
\prod_j\frac{ig}{d^2(p_j,0)-m^2+i\epsilon}, 
\end{equation}
where $i$ counts loops, and $j$ propagators, will always give only a finite contribution.

For the sake of definitness, let us evaluate the leading order correction in the simplest case of a free propagator in $\phi^4$ theory

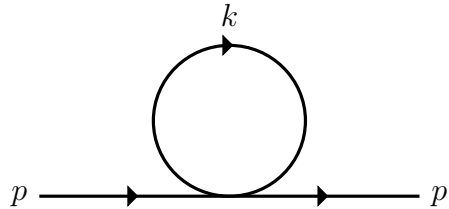
\begin{figure}[h]
\begin{center}
\begin{tikzpicture}[thick]
\begin{scope}[very thick, every node/.style={sloped,allow upside down}]  
\draw   (-2.5,0) -- node{\midarrow} (0,0);
\draw   (-5,0) -- node{\midarrow} (-2.5,0);
\draw (-2.5,1) circle (1);
\draw (-2.5,2) node{\midarrow};
\draw (-5,0) node[anchor= east] {$ p $};
\draw (0,0) node[anchor= west] {$ p $};
\draw (-2.5,2.1) node[anchor= south] {$ k $};
\end{scope} 
\end{tikzpicture}
\end{center}
\end{figure}

\begin{equation}
i\mathcal M=ig\int \frac{d\Omega_k}{(2\pi)^4} \frac{i}{d^2( k,0)-m^2+i\epsilon}. 
\end{equation}
For this purpose orthogonal projection $h=1$ angular coordinates are used, with angles $(\omega,\rho,\theta,\varphi)$. We preform Wick's rotation on the embedding space, as is done in the standard case, to obtain Euclidean integrals. Rotating $\eta_0$ from (\ref{emb}) for $\pi/2$ in the complex plane, the surface becomes a sphere with imaginary radius, and the distance function depends only on the polar angle $\omega$, $d=i\mathcal P\omega$. This gives

\begin{equation}
i\mathcal M=-\frac{ig\mathcal P^4}{8}\int_0^{\pi/2} d\omega \sin^3\omega\frac{i}{\mathcal P^2\omega^2+m^2}. 
\end{equation}
The solution is a complicated expression involving combination of different sine and cosine integrals, so we do not report it here.  For the masses of the order of magnitude of  the Standard Model particles, which is much less than the order of magnitude of $\mathcal P$, usually assumed to be in the range of Planck's mass, one can take the value of the integral at $m/\mathcal P=0$, that is,  

\begin{equation}
i\mathcal M\approx 0.086g\mathcal P^2.
\end{equation}
This value is very large compared to the values of the parameters of the Standard Model, and one still needs to introduce counterterms to renormalize, as in the canonical case. The difference is that in this case one deals only with the finite quantities, thus avoiding the conceptual difficulties related to the dealing with infinities.

Momentum space Feynman rules represent a self-contained set of rules for evaluating invariant amplitudes, whose basic elements are propagators (the inverses of the dispersion relation), energy-momentum conservation law and integrations over the loop momenta. All  this elements have a clear and direct generalization to the space of constant curvature, as defined by our set of axioms and general geometrical properties of such space. Still, one may be interested in deriving the momentum space Feynman rules starting from the spacetime representation, as it is usual in the canonical case. This can be done. We do not report a full derivation here, but only highlight some key technical points.

The configurational space dual to the constant curvature momentum space is discrete, while time remains continuous \cite{mir2,lus1}. The field operators in the space-time picture depend then on three discrete parameters (space) and one continuos (time). The latter point maintains the same structure of the time ordering as in the standard case, as well as derivatives over time. Field commutation relations become

\begin{equation}
[\phi(x,t), \phi(x',t)]=0, \ \ \ \ \
[\phi(x,t), \partial_t\phi(x',t)]=i\delta_{xx'},
\end{equation}
where we use $x$ to represent a triplet of integers, say $(j,l,m)$ and $\delta_{xx'}=\delta_{jj'}\delta_{ll'}\delta_{mm'}$.  Following the same steps as in the standard case, one shows that 

\begin{equation}
(\square+m^2)\langle \Omega| \phi(x,t)\phi(x',t')    |\Omega \rangle=\langle \Omega|(\square+m^2) \phi(x,t)\phi(x',t')    |\Omega \rangle-i\delta_{xx'}\delta(t-t'),
\end{equation}
where the Laplacian $\square$ is represented by differential-finite-difference operator \cite{mir2}, and, along the same lines,

\begin{equation}
(\square_x+m^2)\langle \phi_x\phi_1\dots \phi_n  \rangle=\langle \mathcal L_{int}'[\phi_x]\phi_1\dots\phi_n \rangle-i\sum_j\delta_{xx_j}\delta(t-t_j)\langle  \phi_x\phi_1\cdots \phi_n \rangle,
\end{equation}
where we introduced the standard condensed notation. These are the Schwinger-Dyson equations. They are the same as in the standard case, with the only difference in the Laplacian and the replacement of delta functions with Kronecker deltas. They are solved perturbatively in the same way as in the flat spacetime case. This gives the usual Feynman rules in the position represantation. In particular, the Feynman propagator is

\begin{equation}
D_F(x,y)=\lim_{\epsilon\to0}\int\frac{d\Omega_p}{(2\pi)^4}\frac{i}{d^2(p,0)-m^2+i\epsilon}Y_{x}(p)Y^*_{y}(p)e^{iE(t_x-t_y)},
\end{equation}
where $Y_x(p)$ represents spherical harmonics on the three-sphere. In passing to energy-momentum representation, instead of the integrals over spacetime, we now have sums over space and integrals over time. Importaintly, we have

\begin{equation}
\sum_{x}Y_{x}(p)Y^*_{x}(k)\int dt e^{i(E_p-E_k)t}=\delta(p\ominus k),
\end{equation}
from which the above introduced Feynman rules on the momentum space follow.

The next logical step is to apply the principles introduced here to the case of quantum electrodynamics. The general idea remains the same, but there are two new points to consider: one is the linearisation of the Klein-Gordon equation to obtain Dirac's equation, and the other is the question of the form of the gauge symmetry. We intend to investigate these questions in the future work.


\begin{thebibliography}{9}
\bibitem{stu} E.C.G. Stueckelberg, A. Petermann, Helv. Phys. Acta \textbf{26} 499 (1953)
\bibitem{gm} M. Gell-Mann,  F.E. Low, Phys. Rev. \textbf{95} 1300 (1954)
\bibitem{wein} S. Weinberg, Phys. Rev. 118 838 (1960)
\bibitem{tho} G. 't Hooft, M. Veltman, Nucl. Phys. \textbf{B44} 189 (1972); G. ’t Hooft,  Nucl. Phys. \textbf{B61} 455 (1973) 
\bibitem{wil} K.G. Wilson, Phys. Rev. \textbf{B4} 3174 (1971);  Rev. Mod. Phys. \textbf{47} 773 (1975)
\bibitem{sny} H.S. Snyder, Phys. Rev. \textbf{71} 38 (1947)
\bibitem{gol} Yu. A. Gol'fand, JETP \textbf{10} 356 (1960)
\bibitem{kady1} V.G. Kadyshevsky, Nucl. Phys. B \textbf{141} 477 (1978); V.G. Kadyshevsky, M.D. Mateev, Phys. Lett. B \textbf{106} 139 (1981); A.D. Donkov, V.G. Kadyshevsky, M.D. Mateev, Theor. Math. Phys. \textbf{50} 360 (1982)
\bibitem{kady2} V.G. Kadyshevsky, M.D. Mateev, Il Nuovo Cim. A \textbf{87} 324 (1985); M.V. Chizhov, A.D. Donkov, R.M. Ibadov, V.G. Kadyshevsky, M.D. Mateev, Il Nuovo Cim. A \textbf{87} 350 (1985); M.V. Chizhov, A.D. Donkov, R.M. Ibadov, V.G. Kadyshevsky, M.D. Mateev, Il Nuovo Cim. A \textbf{87} 373 (1985); 
\bibitem{mir1} R.M. Mir-Kasimov, J. Phys. A: Math. Gen. \textbf{24} 4283 (1991)
\bibitem{mir2} R.M. Mir-Kasimov, Phys. Lett. B \textbf{378} 181 (1996)
\bibitem{qg1} T. Yoneya, Mod. Phys. Lett. \textbf{A4} 1587 (1989); N. Seiberg, E. Witten, JHEP \textbf{09} 032 (1999)
\bibitem{qg2} S. Doplicher, K. Fredenhagen, J.E. Roberts, Commun. Math. Phys. \textbf{172} 187
(1995); P.J. Garay, Int. J. Mod. Phys. \textbf{A10} 145 (1995); S. Hossenfelder, Liv. Rev.
Rel. \textbf{16} 2 (2013)
\bibitem{qg3} S. Majid, Foundations of Quantum Group Theory, Cambridge University Press
(2000); A. Connes, Noncommutative Geometry, Academic Press (1994)
\bibitem{bm} M.V. Battisti, S. Meljanac, Phys. Rev. \textbf{D82} 024024 (2010)
\bibitem{gl} F. Girelli, E.R. Livine, JHEP \textbf{1103} 132 (2011)
\bibitem{rv} J.M. Romero, J.D. Vergara, Mod. Phys. Lett. \textbf{A30} 1550155 (2015)
\bibitem{fm} S.A. Franchino-Vinas, S. Mignemi, Nucl. Phys. \textbf{B981} 115871 (2022)
\bibitem{bor} B. Iveti\'c, Phys. Rev. \textbf{D100}, 115047 (2019)
\bibitem{bor2} B. Iveti\'c, arxiv:2209.05351 (2022)
\bibitem{relloc1} G. Amelino-Camelia, L. Freidel, J. Kowalski-Glikman and L. Smolin, Phys.     	     Rev. \textbf{D84} 084010 (2011)
\bibitem{lus1} L. Lu, A. Stern, Nucl. Phys. B \textbf{854} 894 (2012)
\end{thebibliography}
\end{document}